\documentclass[12pt]{article}
\usepackage{amsfonts}
\usepackage{amsmath,mathrsfs,bm,amssymb,color}
\usepackage[width=15cm,height=8.5in]{geometry}
\usepackage{graphicx}

\newtheorem{theorem}{Theorem}[section]
\newtheorem{definition}[theorem]{Definition}
\newtheorem{lemma}[theorem]{Lemma}
\newtheorem{proposition}[theorem]{Proposition}
\newtheorem{corollary}[theorem]{Corollary}
\newtheorem{remark}[theorem]{Remark}
\newtheorem{example}[theorem]{Example}

\providecommand{\bc}[1]{\begin{corollary}\label{#1}}
\providecommand{\ec}{\end{corollary}}
\providecommand{\bp}[1]{\begin{proposition}\label{#1}}
\providecommand{\ep}{\end{proposition}}

\newenvironment{proof}[1][Proof]{{\sc #1.} }{\hfill $\Box$}

\newcommand{\bt}[1]{\begin{theorem}\label{#1}}
\newcommand{\et}{\end{theorem}}
\newcommand{\bl}[1]{\begin{lemma}\label{#1}}
\newcommand{\el}{\end{lemma}}

\newcommand{\br}[1]{\begin{remark}\label{#1}}
\newcommand{\er}{\end{remark}}

\makeatletter
\@addtoreset{equation}{section}
\makeatother

\def\bbbone{{\mathchoice {\rm 1\mskip-4mu l} {\rm 1\mskip-4mu l}
{\rm 1\mskip-4.5mu l} {\rm 1\mskip-5mu l}}}
\def\one{\bbbone}

\providecommand{\eq}[1]{\begin{equation}\label{#1}}
\providecommand{\en}{\end{equation}}

\renewcommand{\d}{\displaystyle}

\providecommand{\qed}{\hfill {$\Box$}
\par\medskip}

\providecommand{\bi}
{\begin{description}}
\providecommand{\ei}{\end{description}}

\providecommand{\RR}{\mathbb{R}}

\providecommand{\kak}[1]{(\ref{#1})}

\providecommand{\F}{{\mathscr{F}}}
\providecommand{\hhh}{{\mathscr{H}}}

\providecommand{\lk}{\left(}
\providecommand{\rk}{\right)}

\providecommand{\T}{{{\mathbb T}^d}}

\providecommand{\s}{\sigma}

\newcommand{\zz}{\mathbb Z^d}

\newcommand{\g}{\frac{1}{d}\sum_{j=1}^d (\cos \theta_j+1)}
\renewcommand{\gg}{\Psi\lk\frac{1}{d}\sum_{j=1}^d (\cos \theta_j+1)\rk}
\newcommand{\h}{H_v}

\makeatletter
\@addtoreset{equation}{section}
\makeatother

\author{
\small Fumio Hiroshima\\
{\small\it Department of Mathematics, Kyushu University}    \\[-0.7ex]
{\small\it  6-10-1 Hakozaki, Fukuoka, 812-8581,  Japan}      \\[-0.7ex]
{\small  {\tt hiroshima@math.kyushu-u.ac.jp}}\\[0.3cm]
\small J\'ozsef L\H{o}rinczi\\
{\it \small School of Mathematics, Loughborough University} \\[-0.7ex]
{\it \small Loughborough LE11 3TU, United Kingdom} \\[-0.7ex]
{\small {\tt  J.Lorinczi@lboro.ac.uk}} \\[-0.7ex]
}
\date{}

\begin{document}
\title
{{\Large \sc The Spectrum of  Non-Local
Discrete Schr\"odinger Operators with a $\delta$-Potential}}
\maketitle
\begin{abstract}
\noindent
The behaviour of the spectral edges (embedded eigenvalues and resonances) is discussed at the two ends
of the continuous spectrum of non-local discrete  Schr\"odinger operators with a $\delta$-potential. These
operators arise by replacing the discrete Laplacian by a strictly increasing $C^1$-function of the discrete
Laplacian. The dependence of the results on this function and the lattice  dimension are explicitly derived.
It is found that while in the case of the discrete  Schr\"odinger operator these behaviours are the same no
matter which end of the continuous spectrum is considered, an asymmetry occurs for the non-local cases. A
classification with respect to the spectral edge behaviour is also offered.
\end{abstract}

\section{Introduction}
\subsection{Non-local discrete  Schr\"odinger operators}

The spectrum of discrete Schr\"odinger operators has been widely studied for both combinatorial Laplacians and
quantum graphs; for some recent summaries see \cite{C97,G01,BCFK06,EKKST08,BK12,P12,KS13} and the references
therein. Specifically, eigenvalue behaviours of discrete Schr\"odinger operators on $l^2(\zz)$ are  discussed
in e.g. \cite{EKW10,bs12, hsss12}. However, for discrete non-local (such as fractional) Schr\"odinger operators
only few results are known.

In this paper we define generalized discrete Schr\"odinger operators which include discrete fractional Schr\"odinger
operators and others whose counterparts on $L^2(\Bbb R^d)$ are currently much studied \cite{BB00,BB09,LM12,K13}.
In \cite{HIL12} we have introduced
a class of generalized Schr\"odinger operators whose kinetic term is given by so called Bernstein functions of the
Laplacian.
These operators are non-local and via a Feynman-Kac representation generate subordinate Brownian motion killed at a
rate given by the potential. Their discrete counterparts studied in this paper also have a probabilistic interpretation
in that they generate continuous time random walks with jumps on $\zz$.

In the present paper we consider a class of Schr\"odinger operators obtained as a strictly increasing $C^1$-function of
the discrete  Laplacian and a $\delta$-potential. This includes, in particular, Bernstein functions (see below) of the 
discrete  Laplacian.
In the presence of a $\delta$-potential the above probabilistic picture then describes free motion with a ``bump" which
can be interpreted as an impurity in space. Our aim here is to investigate the spectrum of such operators, specifically,
embedded eigenvalues and resonances at the edges of the continuous spectrum.

Let $d\geq1$ and $L$ be the standard discrete Laplacian on $l^2(\zz)$ defined by
$$
\d L\psi(x)=\frac{1}{2d}\sum_{|x=y|=1}(\psi(y)-\psi(x)).
$$
Also, let $V(x)=v\delta_{x,0}$ be $\delta$-potential with mass $v$ concentrated on $x=0$, i.e., $V\psi(x)=0$
for $x\not=0$ and $V\psi(x)=v\psi(x)$. Then the operator
\eq{hamiltonian}
h=L+v\delta_{x,0}, \quad v\in\RR
\en
is the discrete  Schr\"odinger operator with $\delta$-potential. In order to define a non-local version of $h$, we use
Fourier transform on $l^2(\zz)$. Let $\T =[-\pi, \pi]^d$ be the $d$-dimensional torus, and set \eq{kuchi}
\hhh=L^2(\T).
\en
The Fourier transform $\F:l^2(\zz)\to \hhh$ is then defined by $\F\psi(\theta)=\sum_{n\in\zz} \psi(n)e^{-in\cdot\theta}$ 
for $\theta=(\theta_1,\ldots,\theta_d)\in \T $. Then the discrete  Laplacian $L$ transforms as
$$
\F L \F^{-1} = \frac{1}{d}\sum_{j=1}^d (\cos \theta_j+1),
$$
i.e., the right hand side above is a multiplication operator on $\hhh$. In this paper we use a non-local
discrete  Laplacian $\Psi(L)$ defined for a suitable function $\Psi$ by applying Fourier transform.

\begin{definition}[\textbf{Non-local discrete  Laplace and Schr\"odinger operators}]\  \\
\noindent
\rm{
For a given $\Psi\in C^1((0,\infty))$ such that $\Psi'(x)>0$, $x\in (0,\infty)$, we define the \emph{non-local
discrete  Laplacian} $\Psi\lk L\rk$ by
$$
\Psi(L)=\F^{-1} \Psi\lk \g\rk  \F.
$$
Also, we call
\eq{10}
h=\Psi (L) + v\delta_{x,0}, \quad v\in\RR,
\en
\emph{non-local discrete  Schr\"odinger operator with $\delta$-potential}.
}
\end{definition}

An example of such a function is $\Psi(u) = u^{\alpha/2}$, $0 < \alpha < 2$, which describes a discrete Laplacian of
fractional order $\alpha/2$. Other specific choices will be given in Example \ref{exs} below. 

Under Fourier transform \kak{10} is mapped into
\eq{11}
\h = \F h \F^{-1}=\Psi\lk \g \rk+v (\Omega,\cdot )_{\hhh} \Omega,
\en
where $\Omega=(2\pi)^{-d/2}\one\in \hhh$.

Since $\s(L)=[0,2]$ and $\Psi$ is strictly increasing, it is immediate that $\s(\Psi(L))=\Psi([0,2])=[\Psi(0),\Psi(2)]$.
In what follows we consider the spectrum of $\h$ instead of $h$. Note that the map $\Phi\mapsto v (\Omega,\Phi) \Omega$ is a
rank-one operator, and thus the continuous spectrum of the rank-one perturbation $\h$ of $L$ is $[\Psi(0),\Psi(2)]$, for
every $v\in\RR$.

\subsection{$\Psi(\ast)$-resonances and $\Psi(\ast)$-modes}
As it will be seen below, for a sufficiently large value of $-v > 0$ there exists an eigenvalue $E_-(v)$ of $\h$
strictly smaller than $\Psi(0)$. Suppose that $E_-(v)\uparrow \Psi(0)$ as $v\uparrow v_0$ with some $v_0\not=0$.
If $\Psi(0)$ is an eigenvalue of $H_{v_0}$, we call the eigenvector associated with $\Psi(0)$ a \emph{$\Psi(0)$-mode}.
If $\Psi(0)$ is not an eigenvalue of $H_{v_0}$, we call it a \emph{$\Psi(0)$-resonance}. Similarly, for a
sufficiently large $v>0$ it will be seen that there exists an eigenvalue $E_+(v)$ strictly larger than  $\Psi(2)$.
Suppose that $E_+(v)\downarrow \Psi(2)$ as $v\downarrow v_2$ with some $v_2\not=0$. If $\Psi(2)$ is an eigenvalue of
$H_{v_2}$, we call the eigenvector associated with $\Psi(2)$ a \emph{$\Psi(2)$-mode}, and a \emph{$\Psi(2)$-resonance}
whenever $\Psi(2)$ is not an eigenvalue of $H_{v_2}$.

For the discrete  Schr\"odinger operator $L+V$ these modes and resonances were studied in e.g. \cite{hsss12}, in
particular, their dependence on the dimension $d$. For $d=1,2$, there is no $0$-mode, $2$-mode, $0$-resonance or
$2$-resonance, for $d=3,4$ there are $0$ and $2$-resonances, and for $d\geq 5$ there are $0$ and $2$-modes. This
shows that the eigenvalue behaviour at both edges ($0$ and $2$) is the same (see Table \ref{N}).
\begin{table}[th]
\begin{center}
\arrayrulewidth=1pt
\def\arraystretch{1.0}
\begin{tabular}{c|c|c|c|c}
\ &
$2$-mode&
$2$-resonance&
$0$-mode&
$0$-resonance\\
\hline
$d=1$&
no&
no&
no&
no\\
\hline
$d=2$&
no&
no&
no&
no\\
\hline
$d=3$&
no&
yes&
no&
yes\\
\hline
$d=4$&
no&
yes&
no&
yes\\
\hline
$d \geq 5$&
yes&
no&
yes&
no\\
 \end{tabular}
\end{center}
\caption{Modes and resonances of $L+V$}
\label{N}
\end{table}%
As it will be seen below, for the case of the fractional Laplacian we have the remarkable fact that the edge behaviours
are in general different at the two sides (see Table \ref{c}).
\begin{table}[ht]
\begin{center}
\arrayrulewidth=1pt
\def\arraystretch{1.0}
\begin{tabular}{c|c|c|c|c}
\ &
$\sqrt 2$-mode&
$\sqrt 2$-resonance&
$0$-mode&
$0$-resonance\\
\hline
$d=1$&
no&
no&
no&
no\\
\hline
$d=2$&
no&
no&
no&
yes\\
\hline
$d=3$&
no&
yes&
yes&
no\\
\hline
$d=4$&
no&
yes&
yes&
no\\
\hline
$d\geq 5$&
yes&
no&
yes&
no
 \end{tabular}
\end{center}
\caption{Modes and resonances of $\sqrt L+V$}
\label{c}
\end{table}
Note that $\s(\sqrt L)=[0,\sqrt 2]$.

\section{Eigenvalues}
\subsection{A criterion for determining the eigenvalues}
Consider the eigenvalue equation
$$
\h\Phi=E\Phi
$$
or, equivalently,
\eq{j2}
E\Phi-\gg\Phi=v(\Omega,\Phi)\Omega.
\en
The following result gives an integral test to spot the eigenvalues of $\h$.
\bl{3}
$E$ is an eigenvalue of $\h$ for a given $v$ if and only if
\eq{4}
\int_{\T }  \frac{1}{\left(E-\gg\right)^2}d\theta < \infty
\en
and
\eq{4'}
\int_{\T }  \frac{1}{E-\gg}d\theta \not= 0.
\en
Furthermore, if $E$ is an eigenvalue of $H_v$, then the coupling constant $v$ satisfies
\eq{5}
v = (2\pi)^d \lk \int_{\T }  \frac{1}{E-\gg}d\theta\rk^{-1}.
\en
\el
\proof
To show the necessity part, suppose that $E$ is an eigenvalue and $\Phi$ an associated eigenvector.
Assuming $(\Omega,\Phi)=0$, we have $(\Omega,\h\Phi)=0$ and thus $(\h \Omega, \Phi)= (\one, \Phi)+v\Phi(0)=v\Phi(0)=0$.
Hence $\h\Phi(x)=L\Phi(x)=E\Phi(x)$, for all $x\in\zz$. Since $L$ has no point spectrum, $\Phi=E\Phi$ is a
contradiction. This gives
$$
(\Omega, \Phi)\not=0 \quad \mbox{and} \quad \d\Phi=\frac{(\Omega,\Phi)}{E-\gg} \in \hhh.
$$
Thus \kak{4} follows, and $(\Omega,\Phi)\not=0$ implies \kak{4'}.

For the sufficiency part, suppose now that \kak{4} and \kak{4'} hold. Define the $L^2(\zz)$-function
$$
\d\Phi=\frac{c f }{E-\gg}
$$
with a chosen $c$. It is straightforward to see that $\Phi$ satisfies $\h\Phi=E\Phi$ whenever for $v$
\eq{4''}
v(2\pi)^{-d}\int_{\T }  \frac{1}{E-\gg}d\theta = 1
\en
holds. By \kak{4'} it follows that there exists $c$ such that \kak{4''} is satisfied, hence $E$ is an
eigenvalue of $\h$.
\qed
In order to investigate $\Psi(\ast)$-resonances and $\Psi(\ast)$-modes we use Lemma \ref{3} and estimate
the two integrals
\begin{align}
&I(E)=\int_{\T }  \frac{1}{|E-\gg|^2}d\theta,\\
&J(E)=\int_{\T }  \frac{1}{E-\gg}d\theta
\end{align}
at the two ends $E=\Psi(\ast)$ of the interval $[\Psi(0),\Psi(2)]$.

\subsection{The location of eigenvalues}
\begin{lemma}
\label{j4}
Let $E\in \RR\setminus [\Psi(0),\Psi(2)]$. Then there exists  $v \neq 0$ such that
 $E$ is
an eigenvalue of $\h$.
\end{lemma}
\proof
In this case it  is easily seen that
$I(E)<\infty$  and $J(E)\not=0$.
Then $E$ is an eigenvalue and $v$ is given by \kak{5}.
\qed
\begin{lemma}
\label{j5}
We have $\s(\h)\cap (\Psi(0),\Psi(2))=\emptyset$, for every $v \neq 0$.
\el
\proof
Due to monotonicity of $\Psi$, there is a unique $x\in (0,2)$ such that $\Psi(E)=\Psi(x)$. Thus
$$
\left|E-\gg\right| \leq
C \left|\frac{1}{d} \sum_{j=1}^d(\cos\theta_j+1-x)\right|
$$
with some $C > 0$. Hence
$$
I(E)\geq \int_\T \frac{1}{|C\frac{1}{d} \sum_{j=1}^d(\cos\theta_j+1-x)|^2 } d\theta.
$$
It is directly seen that the right hand side diverges, and thus the lemma follows.
\qed

Next consider the cases $E=\Psi(2)$ and $E=\Psi(0)$. For a systematic discussion of the eigenvalue behaviour
of $\h$ we introduce the following concept.
\begin{definition}
\label{ab}
{\rm
We say that $\Psi$ is of \emph{$(a,b)$-type} whenever
\begin{align}
&\lim_{x\to0+}\frac{\Psi(x)-\Psi(0)}{x^a}\not= 0,\\
&\lim_{x\to0}\frac{\Psi(2)-\Psi(2-x)}{x^b}\not=0.
\end{align}
}
\end{definition}

\begin{lemma}
\label{12}
Let $\Psi$ be of $(a,b)$-type.
Then we have the following behaviour.
\begin{enumerate}
\item[(1)]
$J(E)\not=0$ for both $E=\Psi(0)$ and  $E = \Psi(2)$.

\item[(2)]
For $E=\Psi(2)$ we have that $I(E)<\infty$ if and only if $d\geq 1+4a$, and $J(E)<\infty$ if and
only if $d\geq 1+2a$.

\item[(3)]
For $E=\Psi(0)$ we have that $I(E)<\infty$ if and only if $d\geq 1+4b$, and $J(E)<\infty$ if and
only if $d\geq 1+2b$.
\end{enumerate}
\el
\proof
Since $\Psi$ is strictly increasing, the first statement follows directly.

Let $\Psi$ be of $(a,b)$-type.
Then we have at $\theta\approx (0,\ldots,0)$,
$$
\Psi(2)-\gg\approx
\lk\frac{1}{2d}\sum_{j=1}^n\theta_j^2\rk^a
$$
and at $\theta\approx(\pi,\ldots,\pi)$,
$$
\gg-\Psi(0)
\approx
\lk \frac{1}{2d}\sum_{j=1}^n(\theta_j-\pi)^2\rk^b.
$$
Hence
\begin{align*}
I(\Psi(2))
\approx \int_{\T }
\frac{1}{\left(\sum_{j=1}^n\theta_j^2\right)^{2a}} d\theta \approx \int_0^1 \frac{r^{d-1}}{r^{4a}} dr,
\end{align*}
and similarly
\begin{align*}
J(\Psi(2))
\approx \int_{\T }
\frac{1}{(\sum_{j=1}^n \theta_j^2)^b} d\theta \approx \int_0^1 \frac{r^{d-1}}{r^{2b}} dr.
\end{align*}
Thus the lemma follows for $E=\Psi(2)$. For the case of $E=\Psi(0)$ the proof is similar.
 
\qed

From these lemmas we can derive the spectral edge behaviour of $\h$. The next theorem is the main result
in this paper.
\bt{main}
Assume that $\Psi$ is of $(a,b)$-type. Let
\begin{align}
&v_2=(2\pi)^d\lk\int_\T\frac{1}{\Psi(2)-\gg}d\theta\rk^{-1}>0,\\
&v_0=(2\pi)^d\lk\int_\T\frac{1}{\Psi(0)-\gg}d\theta\rk^{-1}<0.
\end{align}
The spectral edge behaviour of $\h$ is as follows.
\begin{enumerate}
\item[(1)]
Suppose that $v>0$. Then the following cases occur:
\begin{enumerate}
\item[(i)]
Let $d<1+2b$. Then for all $v>0$ there exists an eigenvalue $E>\Psi(2)$.
\item[(ii)]
Let $1+2b\leq d<1+4b$. Then for $v>v_2$ there exists an eigenvalue $E>\Psi(2)$, while for $v\leq v_2$
there is no eigenvalue.
\item[(iii)]
Let $1+4b\leq d$. Then for $v>v_2$ there exists an eigenvalue $E>\Psi(2)$, for $v=v_2$ the value $E=\Psi(2)$
is an eigenvalue, while $v<v_2$ there is no eigenvalue.
\end{enumerate}
\item[(2)]
Suppose that $v<0$. Then the following cases occur:
\begin{enumerate}
\item[(i)]
Let $d<1+2a$. Then for all $v<0$ there exists an eigenvalue $E<\Psi(0)$.
\item[(ii)]
Let $1+2a\leq d<1+4a$. Then for $v<v_0$ there exists an eigenvalue $E<\Psi(0)$, while for $v\geq v_0$
there is no eigenvalue.
\item[(iii)]
Let $1+4a\leq d$. Then for $v<v_0$ there exists an eigenvalue $E<\Psi(0)$, for $v=v_0$ the value $E=\Psi(0)$
is an eigenvalue, while for $v>v_0$ there is no eigenvalue.
\end{enumerate}
\end{enumerate}
\et
\proof
Consider the case $v>0$ and let $d<1+2b$. Then for all $E>\Psi(2)$ we have $I(E)<\infty$ and $J(E)\not=0$. Thus
there exists $v$ such that  $E$ is an eigenvalue of $\h$.

Let $1+2b\leq  d <1+4b$. Then for all $E>\Psi(2)$ we have that $I(E)<\infty$ and $J(E)\not=0$. Thus $E$ is
an eigenvalue of $\h$. Since $J(E)<\infty$, it follows that $E\downarrow\Psi(2)$ as $v\downarrow v_2>0$. However,
$E=\Psi(2)$ is not an eigenvalue since $I(E)=\infty$.

Let $d\geq 1+4b$. Then for all $E>\Psi(2)$ we have $I(E)<\infty$ and $J(E)\not=0$. Thus  $E$ is an eigenvalue of
$\h$. Since $J(E)<\infty$, we obtain $E\downarrow \Psi(2)$ as $v\downarrow v_2>0$. Since $I(E)<\infty$, we have
that $E=\Psi(2)$ is also an eigenvalue. The cases for $v<0$ can be dealt with similarly.

\qed

\begin{remark}
{\rm Note that in general $-v_0\not=v_2$.}
\end{remark}

\begin{remark}
{\rm
From the above it is seen that the spectral edge behaviour of $\h$ depends on the dimension $d$ as well as on the
parameters $a$ and $b$, and the result is different according to which edge is considered. For a summary see the
tables below.
\begin{table}[h]
\begin{center}
\arrayrulewidth=1pt
\def\arraystretch{1.0}

\begin{tabular}{c|c|c|c}
$v>0$ &
$d<1+2b$ &
$1+2b\leq d <1+4b$&
$1+4b\leq d$\\
\hline
$\Psi(2)$-mode&
no &
no &
yes\\
\hline
$\Psi(2)$-resonance&
no &
yes &
no\\
\end{tabular}
\ \vspace{0.5cm}

\begin{tabular}{c|c|c|c}
$v<0$ &
$d<1+2a$ &
$1+2a\leq d <1+4a$&
$1+4a\leq d$\\
\hline
$\Psi(0)$-mode&
no &
no &
yes\\
\hline
$\Psi(0)$-resonance&
no &
yes &
no\\
 \end{tabular}
\end{center}

\caption{$\Psi(0)$ and $\Psi(2)$-modes and resonances}
\label{a}
\end{table}%
}\end{remark}

It is worthwhile to see the implications more closely for some specific choices of function $\Psi$.
\begin{example}
\label{exs}
\rm{
\hspace{100cm}
\begin{enumerate}
\item[(1)]
\emph{Discrete  Schr\"odinger operator:} Let $\Psi(u)=u$. Then $\Psi$ is of $(1,1)$-type and $\h=L+V$.
See Table \ref{N}.
\item
[(2)]
\emph{Fractional discrete Schr\"odinger operator}: Let $\Psi(u)=u^{\alpha/2}$ for $0<\alpha<2$. Then 
$\Psi$ is of $(\alpha/2,1)$-type and $\h = L^{\alpha/2} + V$.
\item
[(3)]
\emph{Relativistic fractional discrete Schr\"odinger operator}: Let $\Psi(u)=(u+m^{2/\alpha})^{\alpha/2}
-m$ for $0<\alpha<2$ and $m>0$. Then $\Psi$ is of $(1,1)$-type.
\item[(4)]
\emph{Discrete jump-diffusion  operator}: Let $\Psi(u)=u+bu^{\alpha/2}$ with $0<\alpha<2$. Then $\Psi$ 
is of $(\alpha/2,1)$-type.
\item[(5)]
\emph{Rotationally symmetric geometric discrete $\alpha$-stable operator}: Let $\Psi(u)=\log(1+u^{\alpha/2})$ 
for $0<\alpha<2$. Then $\Psi$ is of $(\alpha/2, 1)$-type.
\item[(6)]
\emph{Higher order discrete  Laplacian}: Let $\Psi(u)=u^\beta$ for $\beta>1$. Then $\Psi$ is of 
$(\beta, 1)$-type.
\item[(7)]
\emph{Bernstein functions of the discrete  Laplacian}: Let $\Psi$ be a Bernstein function with vanishing right 
limits, i.e., $\Psi: \Bbb R^+ \to\Bbb R^+$ which can be represented in the form $\d \Psi(u)=bu+\int_0^\infty 
(1-e^{-uy})\nu(dy)$, where $b\geq 0$ and $\nu$ is a L\'evy measure with mass on $(0,\infty)$ satisfying
$\d \int_0^\infty (1\wedge y) \nu(dy) < \infty$. Then it follows that $\d \Psi'(2) =b+\int_0^\infty y e^{-2y}
\nu(dy) \not=0$. Furthermore, since $\Psi$ is concave, we have $a=\alpha/2$ with some $\alpha \geq 2$. Hence
$\Psi$ is of $(\alpha/2,1)$-type with some $0\leq \alpha\leq 2$. Note that the first five examples above
are specific cases of Bernstein functions.
\end{enumerate}
}
\end{example}

\section{A classification of spectral edge behaviour}
The functions $\Psi$ of the discrete  Laplacian can be classified according to the behaviour of the
eigenvalues at the two ends of the interval $[\Psi(0),\Psi(2)]$.
\begin{definition}
{\rm
We call $\Psi$ \emph{normal type} if $\Psi$ is $(1,1)$-type, and \emph{fractional type} if $\Psi$ is
$(\alpha/2,1)$-type with $0<\alpha<2$.
}
\end{definition}
The two types show qualitatively different behaviour and we discuss them separately.

\subsection{Normal type}
Let $\Psi$ be of normal type. In this case the spectral edge behaviour is the same as that of the discrete
Schr\"odinger operator $L+V$. The following result has been obtained in \cite{hsss12}.
\bp{normal}
Let $\Psi$ be normal type. We have the following cases.
\begin{enumerate}
\item[(1)]
Let $d=1$ or $2$. For every $v>0$ there exists an eigenvalue $E>\Psi(2)$, and for every $v<0$ an eigenvalue $E<\Psi(0)$.
\item[(2)]
Let $d=3$ or $4$. If $v>0$, then there exists $v_2>0$ such that for all $v>v_2$ an eigenvalue $E>\Psi(2)$ exists, and
for $v\leq v_2$ no eigenvalue exists. If $v<0$, then there exists $v_0<0$ such that for all $v<v_0$ an eigenvalue
$E<\Psi(0)$ exists, and for $v<v_0$ no eigenvalue exists.
\item[(3)]
Let $d\geq 5$. If $v>0$, then there exists $v_2>0$ such that for all $v>v_2$ an eigenvalue $E>\Psi(2)$ exists, for
$v=v_2$ the value $E=\Psi(2)$ is an eigenvalue, and for $v< v_2$ no eigenvalue exists. If $v<0$, then there exists
$v_0<0$ such that for all $v<v_0$ an eigenvalue $E<\Psi(0)$ exists, for $v=v_0$ the value $E=\Psi(0)$ is an eigenvalue,
and for $v>v_0$ no eigenvalue exists.
\end{enumerate}
\ep
Thus the spectral edge behaviour for positive and negative $v$ is qualitatively the same, and the details only depend on
the dimension $d$.

\subsection{Fractional type}
In the fractional type case we have the following spectral edge behaviour.
\bt{fractional}
Let $\Psi$ be of fractional type. The following cases occur.
\begin{enumerate}
\item[(1)]
If $v>0$, then the spectral edge behaviour is the same as for normal type $\Psi$ with $v>0$.

\item[(2)]
If $v<0$, then we have the following cases:
\begin{enumerate}
\item[(i)]
Let $d<1+\alpha$. Then for every $v<0$ there exists an eigenvalue $E<0$.
\item[(ii)]
Let $1+\alpha\leq d<1+2\alpha$. There exists $v_0<0$ such that for all $v<v_0$ an eigenvalue $E<0$ exists, while for
$v\leq v_0$ no eigenvalue exists.
\item[(iii)]
Let $d\geq 1+2\alpha$. There exists $v_0<0$ such that for all $v<v_0$ an eigenvalue $E<0$ exists, for $v=v_0$ the value
$E=0$ is an eigenvalue, and for $v> v_0$ no eigenvalue exists.
\end{enumerate}
\end{enumerate}
\et
In the fractional case it is seen that the edge behaviour for positive and negative $v$ are in general different from
each other, in contrast with the normal type case.

\subsubsection{The case of $\alpha=1$}
For $\alpha=1$ the spectral edge behaviour of $\h=\sqrt L+V$ is displayed for dimensions $d=1,...,4$ and $d\geq 5$ in
Table \ref{c}.
For dimensions $d=2,3,4$ the edge behaviours at $0$ and $\sqrt 2$ are
again different. We have displayed the specific situations in Figures 1-4 below, where $\oplus$ denotes a
resonance, $\bullet$ an eigenvalue, and $\times$ denotes a value which is not an eigenvalue.

\begin{figure}[t]
\begin{center}
\begin{tabular}{c}

\begin{minipage}{0.4\hsize}
\begin{center}
\includegraphics[width=170pt]{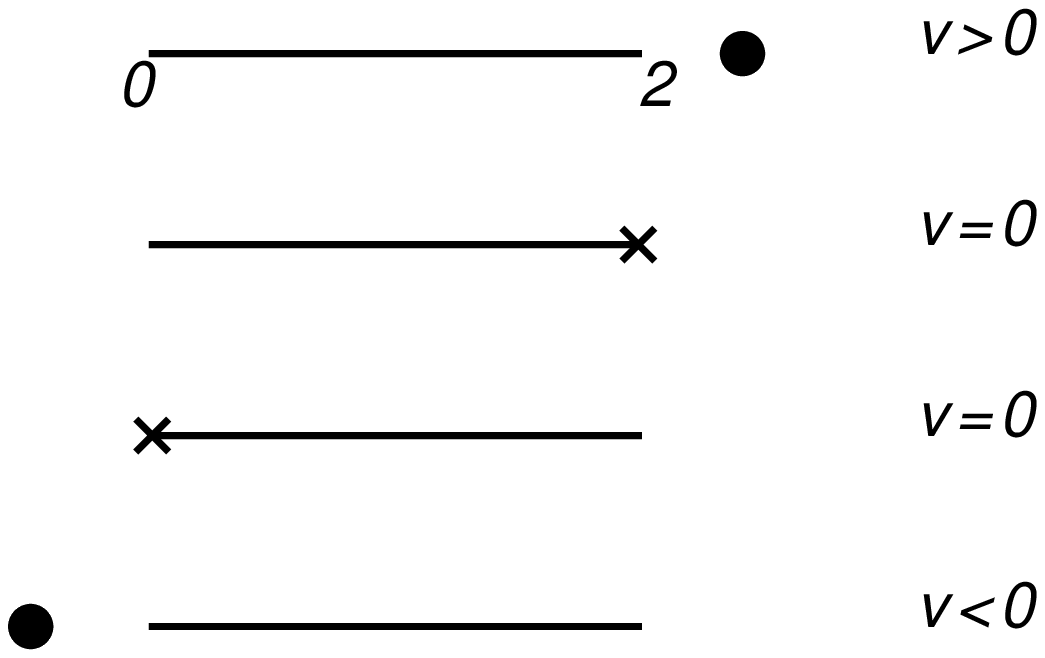}
\caption{$\alpha=1$ and $d=1$}
\end{center}
\end{minipage}
\quad \quad \quad
\quad \quad \quad

\begin{minipage}{0.4\hsize}
        \begin{center}
        \includegraphics[width=170pt]{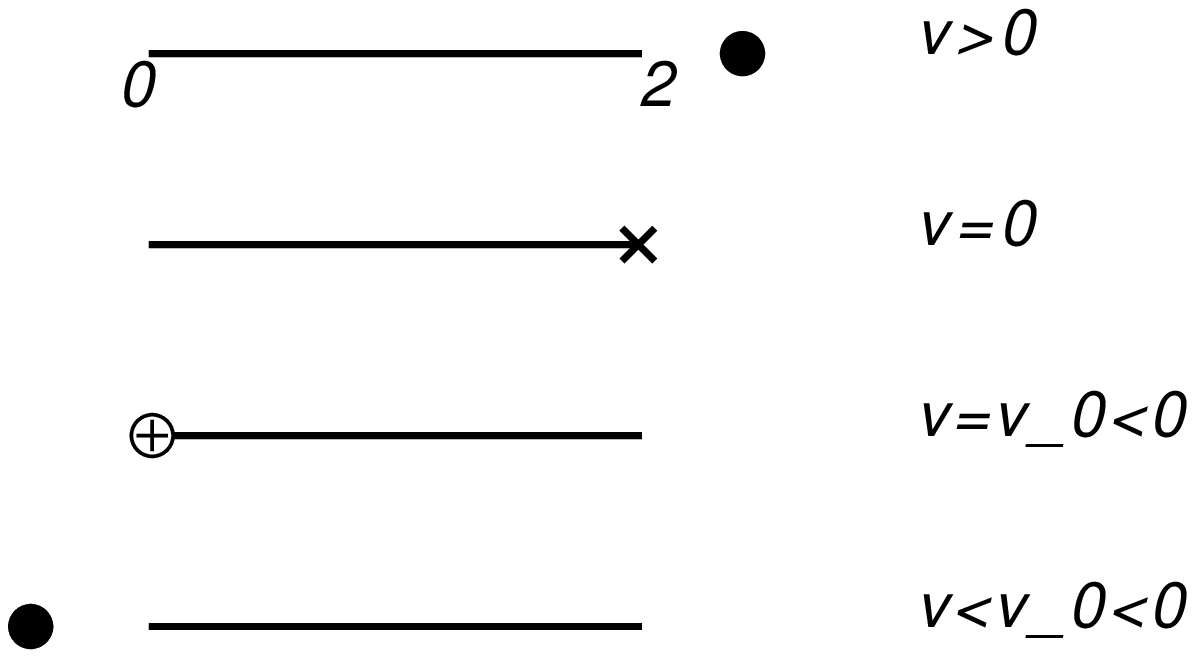}

          \caption{$\alpha=1$ and $d=2$
          }

                  \end{center}
      \end{minipage}

    \end{tabular}
  \end{center}
  \begin{center}
    \begin{tabular}{c}

      \begin{minipage}{0.4\hsize}
        \begin{center}

  \includegraphics[width=170pt]{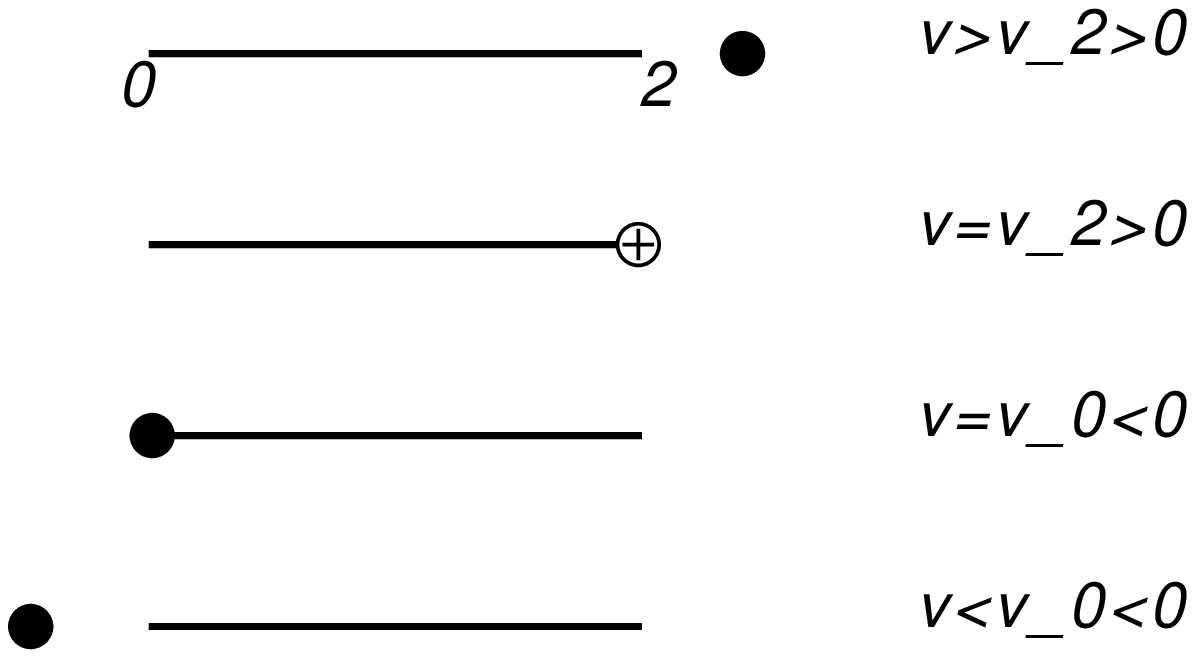}
         \caption{$\alpha=1$ and $d=3,4$}

                  \end{center}
      \end{minipage}
\quad \quad \quad
\quad \quad \quad
      \begin{minipage}{0.4\hsize}
        \begin{center}

  \includegraphics[width=170pt]{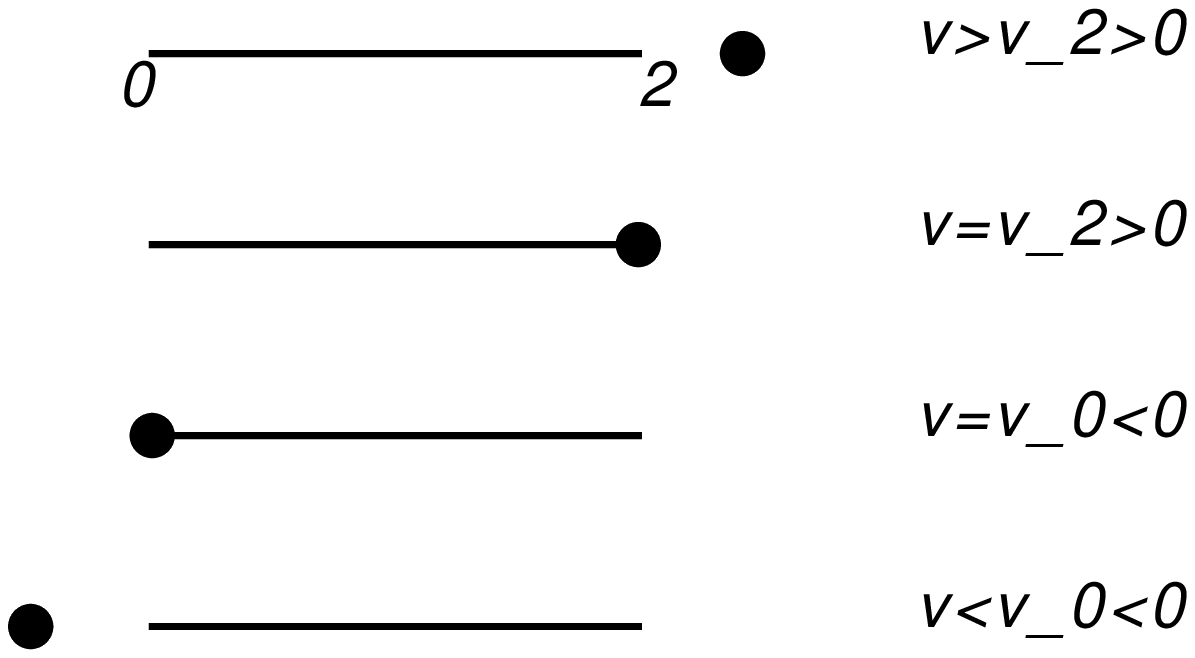}
          \caption{$\alpha=1$ and $d\geq 5$}

                  \end{center}
      \end{minipage}

    \end{tabular}
  \end{center}
\end{figure}

\subsubsection{Massless and massive cases}
Consider the Bernstein function $\Psi(u)=\sqrt {u+m^2}-m$ with $m\geq0$. This allows to define the relativistic
discrete  Schr\"odinger operator $\sqrt{L+m^2}-m+V$. Then it follows that $\Psi(u)$ is $(1,1)$-type for $m>0$,
and $(1/2,1)$-type for $m=0$. In particular, the edge behaviours of $\sqrt{L}+V$ and $\sqrt{L+m^2}-m+V$ are
different. More generally, consider the Bernstein function $\Psi(u)=(u+m^{2/\alpha})^{\alpha/2}-m$, with
$0<\alpha<2$ and $m \geq 0$. This defines the relativistic rotationally symmetric $\alpha$-stable operator
$(L+m^{2/\alpha})^{\alpha/2}-m$. We conclude that $\Psi(u)$ is of $(1,1)$-type for $m>0$ but of
$(\alpha/2,1)$-type for $m=0$. Thus the edge behaviours of $(L+m^{2/\alpha})^{\alpha/2}-m+V$ and $(L)^{\alpha/2}
+V$ are different.

\bigskip
\noindent
\textbf{Acknowledgments.} FH is financially supported by Grant-in-Aid for Science Research (B) 20340032 from JSPS.
JL thanks Institut Mittag-Leffler, Stockholm, for the opportunity to organise the research-in-peace workshop
``Lieb-Thirring-type bounds for a class of Feller processes perturbed by a potential" during the period 25
July~--~9 August 2013. JL also thanks London Mathematical Society for a travel grant to this workshop. FH thanks the
invitation to this workshop, and we both express our gratitude to IML for the kind hospitality and inspiring work
environment, where most of this paper has been prepared.

\end{document}